\documentclass[twocolumn,superscriptaddress,preprintnumbers,prl,amsmath,amssymb]{revtex4}
\usepackage{graphicx}
\usepackage{bm}
\usepackage{epsfig}

\newcommand{\Ba}{BaFe$_2$As$_2$}
\newcommand{\Sr}{SrFe$_2$As$_2$}
\newcommand{\Ca}{CaFe$_2$As$_2$}
\newcommand{\BaSr}{Ba$_{1-x}$Sr$_x$Fe$_2$As$_2$}
\newcommand{\SrCa}{Sr$_{1-y}$Ca$_y$Fe$_2$As$_2$}
\newcommand{\BaSrCa}{(Ba,Sr,Ca)Fe$_2$As$_2$}

\newcommand{\tn}{$T_0$}

\newcommand{\ie}{{\it i.e.}}
\newcommand{\eg}{{\it e.g.}}
\newcommand{\etal}{{\it et al.}}

\begin{document}


\title{Tuning magnetism in FeAs-based materials via tetrahedral structure}


\author{K.~Kirshenbaum}
\author{N.~P.~Butch}
\author{S.~R.~Saha}
\affiliation{Center for Nanophysics and Advanced Materials, Department of Physics, University of Maryland, College Park, MD 20742}

\author{P. Y. Zavalij}
\affiliation{Department of Chemistry and Biochemistry, University of Maryland, College Park, MD 20742}

\author{B. G. Ueland}
\author{J. W. Lynn}
\affiliation{NIST Center for Neutron Research, National Institute of Standards and Technology, Gaithersburg, MD 20899}

\author{J.~Paglione}
\email{paglione@umd.edu}
\affiliation{Center for Nanophysics and Advanced Materials, Department of Physics, University of Maryland, College Park, MD 20742}

\date{\today}


\begin{abstract}

Resistivity, magnetic susceptibility, neutron scattering and x-ray crystallography measurements were used to study the evolution of magnetic order and crystallographic structure in single-crystal samples of the Ba$_{1-x}$Sr$_x$Fe$_2$As$_2$ and Sr$_{1-y}$Ca$_y$Fe$_2$As$_2$ series. A non-monotonic dependence of the magnetic ordering temperature $T_0$ on chemical pressure is compared to the progression of the antiferromagnetic staggered moment, characteristics of the ordering transition and structural parameters to reveal a distinct relationship between the magnetic energy scale and the tetrahedral bond angle, even far above $T_0$. In Sr$_{1-y}$Ca$_y$Fe$_2$As$_2$, an abrupt drop in $T_0$ precisely at the Ca concentration where the tetrahedral structure approaches the ideal geometry indicates a strong coupling between the orbital bonding structure and the stabilization of magnetic order, providing strong constraints on the nature of magnetism in the iron-arsenide superconducting parent compounds.

\end{abstract}


\maketitle


A key question in the quest to understand the mechanism behind high-temperature superconductivity in iron-pnictide and iron-chalcogenide based materials involves understanding the roles of structure and magnetism, and the interplay between them \cite{review}.
For magnetism, there is an ongoing debate in classifying the nature of spin interactions: while a local-moment Heisenberg exchange interaction can be used to describe high-energy spin waves \cite{Zhao_Sr,Zhao_Ca}, unphysical anisotropic interactions as well as small magnetic moment size \cite{Yang,Diallo} point to a more complicated scenario involving itinerant magnetism, frustration \cite{Si_corr,Rodriguez,Nevidomskyy}, orbital order \cite{Chen} or a more complex scenario \cite{Zhao_Ca,Johannes,Tesanovic}. 
The nature and role of bonding between iron and arsenic is widely thought to hold the key information behind the intriguing physical properties of iron-based superconductors, with strong covalency \cite{Yang} and sensitivity to the degree of As-Fe hybridization \cite{Yildirim}, most pronounced in the strong lattice collapse observed in \Ca\ under pressure \cite{Kreyssig} and chemical substitution \cite{Saha_CaR}. 

The internal FeAs$_4$ structure, in particular the specific shape of the iron-pnictide tetrahedron, was suggested early on to play a key role in driving structural and magnetic transitions in the iron-pnictide materials \cite{McQueen1,ZhaoDai}, and continues to appear important to superconducting properties. In particular, the correlation between ideal tetrahedral bond angle and optimal superconducting critical temperature of the ferropnicitides \cite{Lee,ZhaoDai,Kreyssig,Horigane} remains as an elusive property of obvious importance.
Here we demonstrate that an intimate relationship exists in the (Ba,Sr,Ca)Fe$_2$As$_2$ series of compounds between the tetrahedral substructure and the stabilization of long-range magnetic order, with an intriguing evolution of the magnetic ordering temperature as a function of alkaline earth substitution that is dictated by the tetrahedral structure. Evidenced by a correspondence between abrupt features in both the tetrahedral bond angle and magnetic ordering temperature as a function of chemical pressure in the \SrCa\ series, this suggests a direct relationship between structural tuning and the magnetic energy scale of the iron-based superconducting materials.

Single-crystal samples of \BaSr\ and \SrCa\ were grown using the FeAs self-flux method \cite{Saha_Sr}. 
Crystal structures were refined (SHELXL-97 package) using the I4/mmm space group against 113 and 106 independent reflections measured at 250 K with a Bruker Smart Apex2 diffractometer and corrected for absorption using the integration method based on face indexing (SADABS software). Substitution concentrations $x$ and $y$ were refined to within $\pm 0.01$ of values quoted below, with final R-factors in the range 1-2\%.
Chemical analysis was obtained via both energy-dispersive and wavelength-dispersive x-ray spectroscopy, showing 1:2:2 stoichiometry and Ca concentration values reported herein.
Resistivity $\rho$ was measured with the standard four-probe ac method and magnetic susceptibility $\chi$ was measured in a SQUID magnetometer.
Neutron scattering experiments were performed on single-crystal samples using the BT9 triple axis spectrometer at the NIST Center for Neutron Research.  
Diffraction measurements were made using the (002) reflection from a pyrolitic graphite monochromator, which yielded a fixed incident wavelength of 2.359~\AA.  
For measurements of the structural transition, the diffracted beam was analyzed using the (002) reflection of a pyrolitic graphite crystal, and tight collimations of 10'-M-10.7'-S-10'-A-10' were used, where M, S, and A are the monochromator, sample, and analyzer, respectively.  Magnetic moments and order parameter temperature dependence were determined using two-axis mode with 40'-M-47'-S-42' collimation.

\begin{figure}[!t]\centering
  \resizebox{7cm}{!}{
  \includegraphics{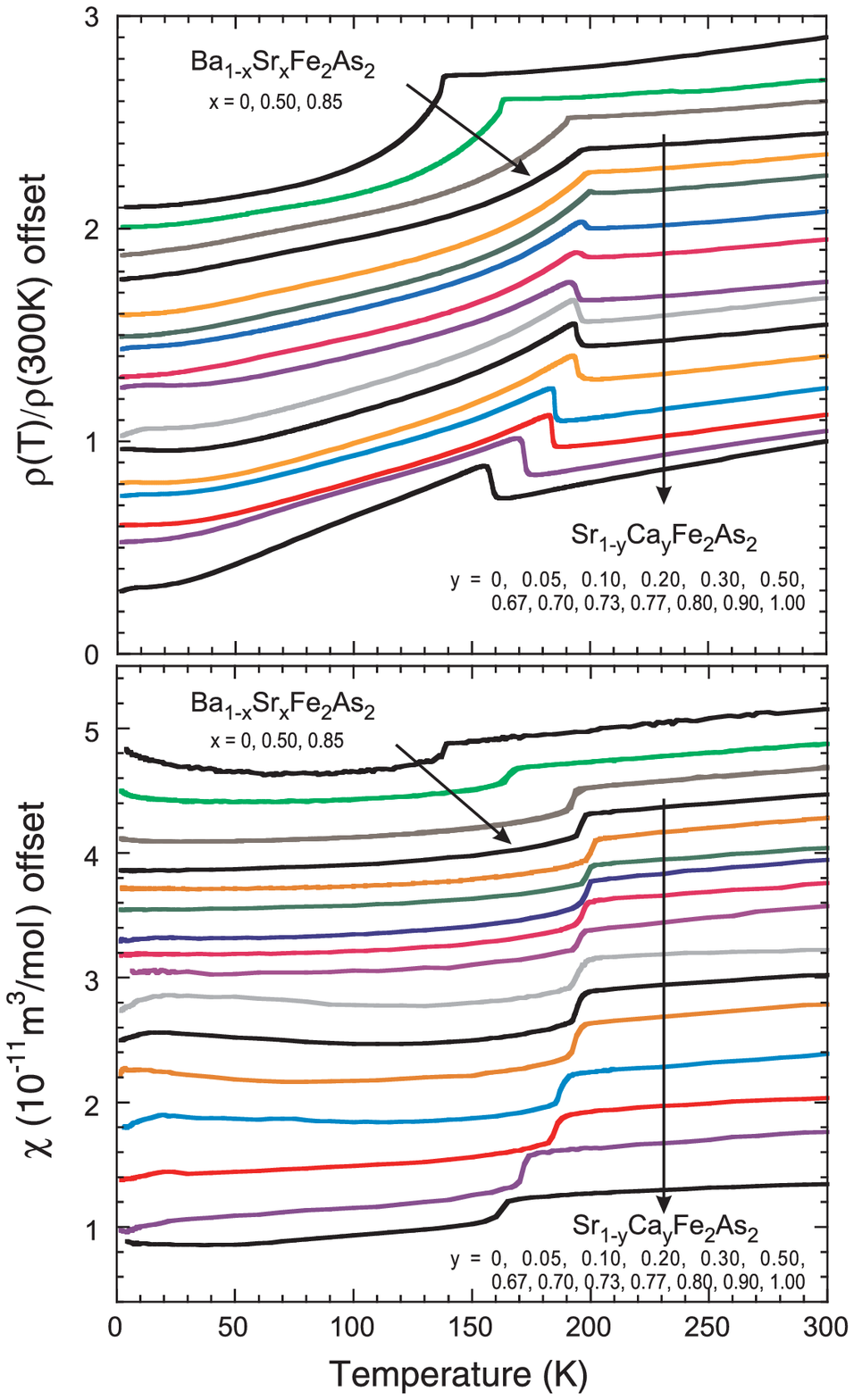}}               				
  \caption{\label{rhochi} (a) Evolution of electrical resistivity of single-crystal samples of \BaSr\ and \SrCa\ with alkaline earth substitution, normalized to 300~K and offset from $y$=1 for clarity.
  (b) Magnetic susceptibility of \BaSr\ and \SrCa\ crystals measured in 10~mT field oriented along basal plane direction, offset from $y$=1 for clarity.}
\end{figure}

As shown in previous work \cite{SCES,Wang}, the unit cell volume of the \BaSrCa\ solid solution series steadily decreases in a linear manner with $x$ and $y$ in both \BaSr\ and \SrCa, respectively, as expected by Vegard's law. Furthermore, it decreases as a function of substitution continuously and at the same rate for both series, showing that the choice of alkaline earth substitution  provides a tunable and uniform chemical pressure effect.
The antiferromagnetic transition \tn, on the other hand, does not follow a monotonic evolution with unit cell volume, but rather finds a maximum in \Sr\ near 200~K with lower values of 135~K and 160~K in \Ba\ and \Ca, respectively. With magnetic order in these materials likely being at least partly itinerant in nature \cite{LynnDai}, the value of \tn\ will depend on details of the electronic structure (\eg, the antiferromagnetic (AFM) nesting condition for a spin-density wave model) and hence may indirectly depend on unit cell parameters. If this is so, abrupt changes in \tn\ with alkaline substitution are not expected as long as the change in chemical pressure is uniform and monotonic.

\begin{figure}[!t]\centering
   \resizebox{8cm}{!}{
   \includegraphics{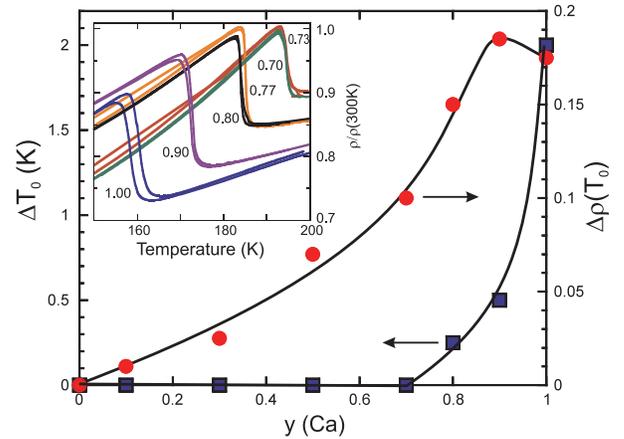}}               				
   \caption{\label{hyst} Characterization of the first-order antiferromagnetic transition observed in electrical resistivity measurements of \SrCa, showing the temperature width of thermal hysteresis $\Delta T_0$ (squares) observed in resistivity data (inset), as well as the relative magnitude of the jump in resistivity $\Delta\rho(T_0)$ (circles) relative to $\rho$(300~K).}
\end{figure}

As shown in Fig.~\ref{rhochi}, the substitution of Sr into \BaSr\ and Ca into \SrCa\ appears to induce very little qualitative change in either $\rho(T)$ or $\chi(T)$ as a function of substitution. An abrupt feature appears in both $\rho(T)$ and $\chi(T)$ at the magnetostructural transition \tn, which climbs to 200~K in the \BaSr\ series and then remains ominously fixed until very high Ca concentrations in the \SrCa\ series, where it begins to decrease in temperature toward 165~K in \Ca. 
No obvious change is observed in the qualitative shape of the \tn\ transition in $\chi(T)$ data, which shows a step-like feature through the entire range of substitutions that remains almost identical in width and height. In contrast, the transport feature associated with \tn\ displays a continuous evolution from a simple but sharp shoulder in \BaSr\ toward a pronounced step-like feature in \Ca. 

The step-like transition in $\rho(T)$ that grows with increasing Ca concentration is consistent with the evolution of the transition from `weakly' to `strongly' first-order in character.  {\it Ab initio} calculations suggest that this is due to a change in Fermi surface nesting features with lattice density \cite{Zhang}. 
Interestingly, both continuous and abrupt features associated with this evolution are shown by the progression of features in $\rho(T_0)$. As shown in Fig.~\ref{hyst}, the emergence of the step feature at \tn\ appears almost immediately upon Ca substitution and continuously grows in size toward the pure Ca end, while a pronounced hysteresis between warming and cooling curves appears only at 70\% Ca, increasing in temperature width very rapidly toward 100\% Ca. This abrupt appearance of strong first-order character is coincident with a sudden decrease in \tn\ with increasing Ca content near a critical concentration $y_c$=0.70.

\begin{figure}[!t] \centering
  \resizebox{7cm}{!}{
  \includegraphics{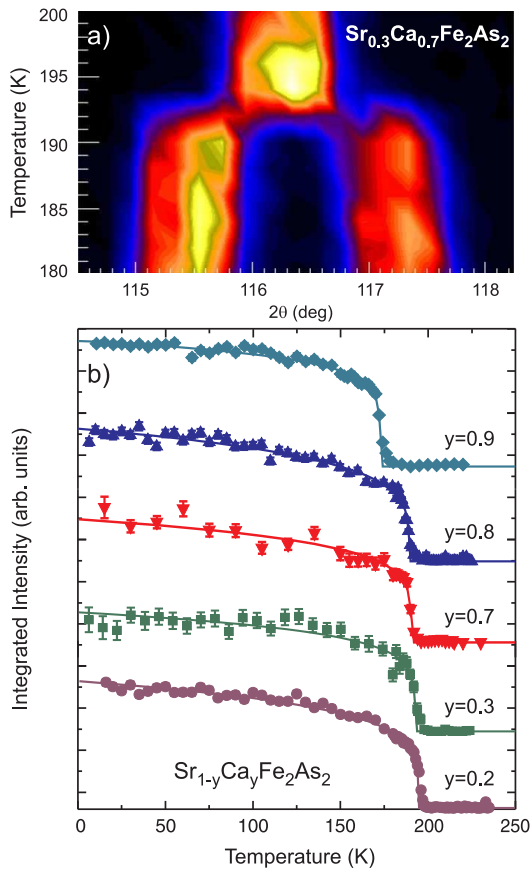}}
  \caption{\label{neutron} a) Evolution of the (220) structural peak through the magnetic/structural transition in Sr$_{0.3}$Ca$_{0.7}$Fe$_2$As$_2$, demonstrating the abrupt onset of orthorhombic splitting at $T_0=193$~K. b) Magnetic order parameter of \SrCa\ single-crystal samples (errors represent one standard deviation) obtained from the (103) magnetic peak. Lines are fits as discussed in text.}
\end{figure}

In order to probe the nature of the magnetic transition through this concentration, elastic neutron scattering experiments were performed on several single-crystal samples of \SrCa. 
Fig.~\ref{neutron}a presents an image plot demonstrating the tetragonal to orthorhombic transition in Sr$_{0.3}$Ca$_{0.7}$Fe$_2$As$_2$, generated from theta two-theta scans of the tetragonal phase's (220) Bragg peak, which splits abruptly at \tn\ into the (400) and (040) Bragg peaks of the orthorhombic phase.
Shown in Fig.~\ref{neutron}b, the magnetic order parameter obtained from the (103) magnetic Bragg peak remains surprisingly similar across the Sr-Ca series, with an abrupt onset at \tn\ consistent with a first-order transition as evidenced by a lack of critical scattering both above and below \tn\ in both \Sr\ \cite{Zhao_Sr} and \Ca\ \cite{Zhao_Ca} end members.

\begin{figure}[!t] \centering
  \resizebox{8cm}{!}{
  \includegraphics[width=8cm]{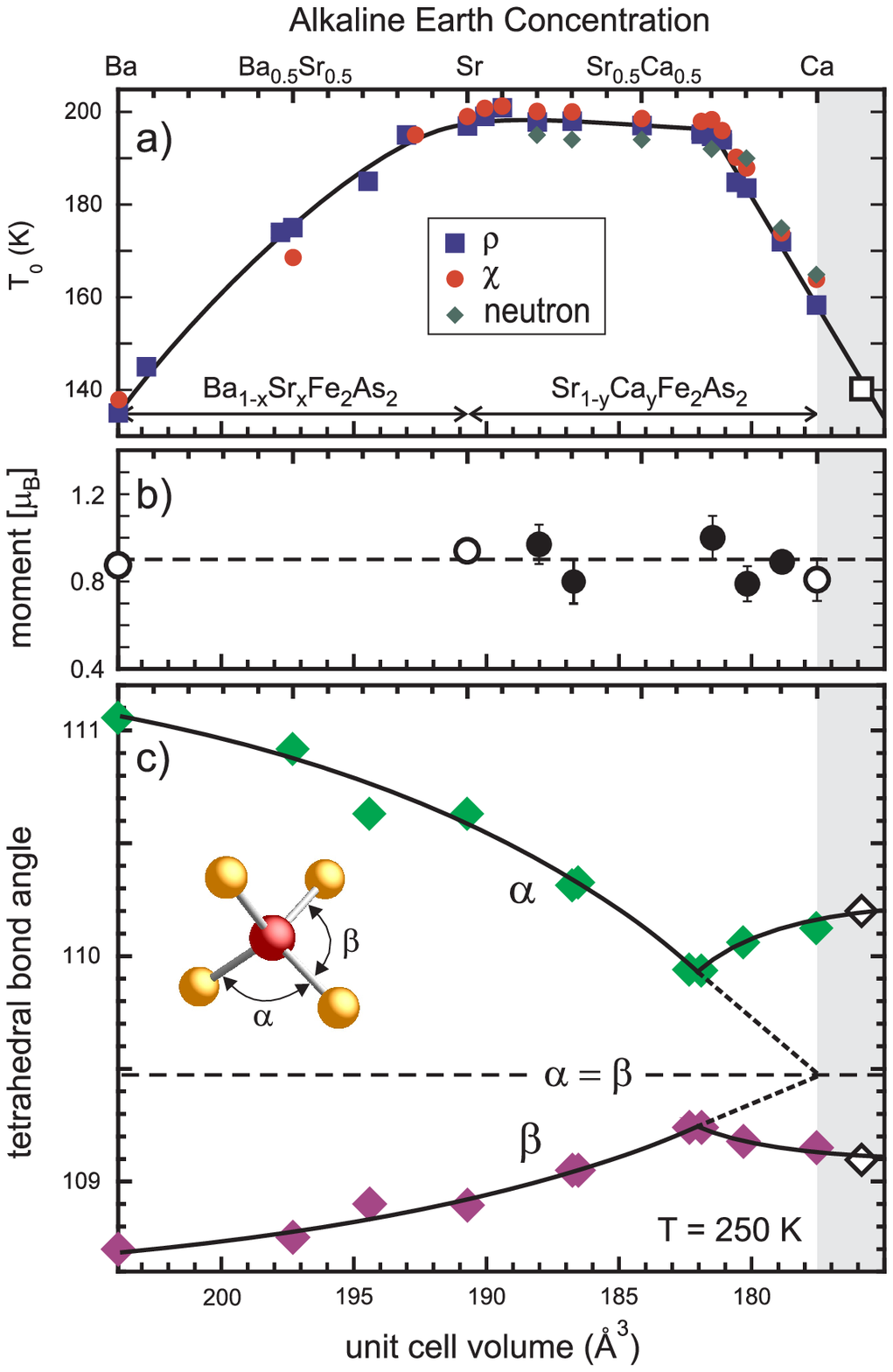}}
  \caption{\label{bondangle} Evolution of structural and magnetic properties of \BaSr\ and \SrCa\ as a function of solution concentration (top axis) or experimental unit cell volume (bottom axis). (a) Magnetic transition \tn\ identified by resistivity (squares), susceptibility (circles) and neutron scattering (diamonds); (b) staggered moment of the antiferromagnetic phase (data for \Ba, \Sr\ and \Ca\ (open circles) obtained from Refs.~\onlinecite{Huang},~\onlinecite{Zhao} and ~\onlinecite{Goldman_Ca}, respectively), with dashed line indicating a moment size of 0.9~$\mu_B$; (c) tetrahedral bond angles $\alpha$ and $\beta$ as identified in the graphic. Dashed line indicates the ideal tetrahedron geometry where $\alpha$=$\beta$=109.47$^{\circ}$. The shaded regions indicate transition \cite{Goldman} and bond angle \cite{Kreyssig} data for \Ca\ at 0.23~GPa.}
\end{figure}

The temperature dependence of the ordered moment does not visibly change through the entire range of Ca concentrations. Fitting to a mean-field/power law form (shown by solid lines) yields an exponent $\simeq 0.20$ (constant within error for all concentrations studied) that lies between those reported for \Ba\ ($\simeq 0.10$) and several doped systems with larger exponents ($\simeq 0.25$) \cite{Wilson}, but is obviously strongly affected by the presence of a first-order jump in the order parameter.
Similar to transport and susceptibility data above, the AFM ordering transition is stagnant with increasing Ca concentration until high concentrations, where it begins to drop toward the \Ca\ end member value. Surprisingly, aside from the abrupt decrease in \tn\ above $y_c=0.70$, there is no change in behavior of the magnetic order parameter, either qualitatively or quantitatively, through this critical concentration. This includes the size of the ordered moment, which remains at $0.9~\mu_B$ across the entire \BaSrCa\ series  to within experimental error (Fig.~\ref{bondangle}b), as well as the order parameter temperature dependence itself (Fig.~\ref{neutron}b). 
Together with the featureless evolution of the character of the transition in $\chi(T)$ data and the transformation observed in $\rho(T)$, this suggests that the the \tn\ transition has more impact on the charge carriers than the magnetic response, consistent with an itinerant (\ie, spin-density wave) form of magnetic order. 

The lack of correspondence between \tn\ and the size of the ordered moment puts strong constraints on the nature of the magnetic interaction.  In a simple model of AFM, the N\'eel temperature is proportional to both the ordered (staggered) moment and the exchange coupling. In contrast to the direct proportionality between \tn\ and the ordered moment observed in both Co- \cite{Fernandes} and Ru-doped \cite{Kim}  \Ba\ as well as P-doped
CeFeAsO \cite{delaCruz}, the absence of any correlation between the ordered moment size and \tn\ in \SrCa\ suggests that the variation of \tn\ in the \BaSrCa\ series results primarily from the tuning of exchange.
Lacking any direct manipulation of electronic structure in this series (\eg, from charge doping), structural tuning must play a direct role in setting the magnetic energy scale.

Using refinements of single-crystal x-ray data for \BaSr\ and \SrCa\ series obtained at 250~K, the internal structure of the unit cell is plotted in the form of As-Fe-As tetrahedral bond angles $\alpha$ and $\beta$, together with the evolution of \tn\ and the ordered moment in Fig.~\ref{bondangle}. While showing a general procession indicative of a greater sensitivity to $a$-axis reduction than the $c$-axis decrease across the series \cite{SCES}, a non-monotonic inflection in both angles appears to coincide precisely with the critical concentration $y_c$. This is clear evidence for a direct correlation between the magnetic energy scale and details of the internal crystal structure involving the FeAs layer. Moreover, with signatures of the mechanism that controls the energy scale for magnetic ordering occurring well above \tn\ (\ie, at 250~K), it appears that the crystal structure plays a precursor role in determining the magnetic energy scale.

Sensitivity of magnetic order to fine-tuning of the lattice structure is surprising in light of (a) the strong first-order nature of the magnetic transition, and (b) the widely held view that the structural transition that accompanies \tn\ is driven by magnetic interactions (and not vise-versa) \cite{review}. 
However, with critical scattering persisting up to temperatures high above \tn\ \cite{Wilson2}, it is possible that magnetic interactions do play a primary role. 
But because there is no clear indication of local-moment type order (\eg, no direct relationship between ordering and any structural bond length, \ie, tuning $J$), it is tempting to assign the observed coupling between magnetic ordering and structure to details involving the electronic structure, in particular the nesting condition that is thought to favor magnetic ordering in the parent compounds and to play a vital role in optimizing superconductivity \cite{Usui}. Measurements probing this idea, such as photoemission and quantum oscillations experiments, are thus a promising route to elucidating the tie between magnetic and structural features of the iron-pnictides.

Finally, note that while the tetrahedral bond angle never drops below 110$^{\circ}$ through the \BaSrCa\ series, it comes very close to the value of 109.47$^{\circ}$ expected for an ideal tetrahedral geometry at the critical concentration $y_c$. This particular concentration is ideal for further study of the relationship between structure and superconductivity that has been previously highlighted. In particular, an extrapolation of $\alpha$ to $x=1$ extends to a value very close to 109.47$^{\circ}$, suggesting that the application of pressure may drive a sample with $y=y_c$ closer to this value \cite{Jeffries}. What causes this reversal of structural evolution close to the \Ca\ end-member is currently not understood, but the possibility of a change in $c$-axis coupling at a particular unit cell dimension may coincide with the abrupt features observed at $y_c$.


The authors acknowledge useful discussions with I.I.~Mazin and M.A. Green. 
This work was supported by AFOSR-MURI Grant No. FA9550-09-1-0603 and
NSF-CAREER Grant No. DMR-0952716.


\end{document}